# Voltage-controlled topological spin textures in the monolayer limit

Yangliu Wu[1,2,7], Bo Peng[1,2,7] ✉, Zhaozhuo Zeng[3,7], Chendi Yang[4], Haipeng Lu[1,2], Peiheng Zhou[1,2], Jianliang Xie[1,2], Difei Liang[1,2], Linbo Zhang[1,2], Peng Yan[3] ✉, Haizhong Guo[5,6] ✉, Renchao Che[4] ✉ and Longjiang Deng[1,2] ✉

[1]National Engineering Research Center of Electromagnetic Radiation Control Materials, School of Electronic Science and Engineering, University of Electronic Science and Technology of China, Chengdu 611731, China

[2]Key Laboratory of Multi Spectral Absorbing Materials and Structures of Ministry of Education, School of Electronic Science and Engineering, University of Electronic Science and Technology of China, Chengdu 611731, China

[3]School of Physics and State Key Laboratory of Electronic Thin Films and Integrated Devices, University of Electronic Science and Technology of China, Chengdu, 610054, China

[4]Laboratory of Advanced Materials, Department of Materials Science, Collaborative Innovation Center of Chemistry for Energy Materials(iChEM), Fudan University, Shanghai 200433, China

[5]School of Physics, Zhengzhou University, Zhengzhou 450052, P. R. China

[6]Institute of Quantum Materials and Physics, Henan Academy of Sciences, Zhengzhou 450046, China

[7]These authors contributed equally: Yangliu Wu, Bo Peng, Zhaozhuo Zeng

✉ To whom correspondence should be addressed. Email address: bo_peng@uestc.edu.cn; denglj@uestc.edu.cn; rcche@fudan.edu.cn; yan@uestc.edu.cn; hguo@zzu.edu.cn

**Abstract:** The physics of phase transitions in low-dimensional systems has long been a subject of significant research interest. Recently, the discovery of long-range magnetic order in the strict two-dimensional (2D) limit[1, 2] has circumvented the constraints imposed by the Mermin-Wagner theorem[3], rapidly emerging as a focal point of research. However, the demonstration of a non-trivial topological spin textures in 2D limit has remained elusive. Here, we demonstrate the out-of-plane electric field breaks inversion symmetry while simultaneously modulating the electronic band structure, enabling electrically tunable spin-orbit interaction (SOI) for creation and manipulation of topological spin textures in monolayer $CrI_3$. The realization of ideal 2D topological spin states may offer not only an experimental testbed for probing the Berezinskii–Kosterlitz–Thouless (BKT) mechanism[4, 5], but also potential insights into unresolved quantum phenomena including superconductivity and superfluidity[6, 7]. Moreover, voltage-controlled SOI offers a novel pathway to engineer two-dimensional spin textures with tailored symmetries and topologies, while opening avenues for skyrmion-based next-generation information technologies[8, 9].

The pursuit of low-dimensional magnetic materials and their unique topological spin textures dates back to the last century. Early theoretical studies recognized that while the Mermin-Wagner theorem prohibits long-range magnetic order in low-dimensional systems ($d < 3$) with continuous symmetry[3], this limitation could be overcome through substantial magnetic anisotropy. This is evidenced by recent realizations of Ising like 2D magnets[1, 2], such as monolayer ferromagnetic $CrI_3$ and $Cr_2Ge_2Te_6$, which have since sparked extensive research interest in this field. The BKT theory predicts the stabilization of topological spin textures in rigorously two-dimensional systems[4, 5]. However, conclusive experimental verification of topological spin textures in atomically thin monolayers remains a significant challenge. The topological spin textures in real-space and band structures of non-trivial topology in $k$-space (momentum space), have attracted enormous attention due to their elegant Berry curvature physics[10-16]. Topologically protected, particle-like spin textures—particularly skyrmions, which exhibit robust stability and controlled motion under electrical stimuli (e.g., applied currents)—represent promising candidates for information carriers in next-generation memory and logic devices[17-20]. Among the key challenges in developing future energy-efficient devices is achieving magnetic state switching via electric fields[21, 22]. Despite early demonstrations of electric-field switching of skyrmions through STM tips[23, 24], reports on their translation into practical tunnel junction nanodevices remain scarce[25].

Here we demonstrate the electrical creation of the topological spin textures in monolayer $CrI_3$, which we detect through the unconventional topological magnetic circular dichroism. The gate-controlled modulation of both Dzyaloshinskii-Moriya interaction (DMI) and magnetic anisotropy energy (MAE) drives a reversible phase transition from ferromagnetic order to stabilized topological spin states. Electrostatic gating enables continuous modulation of the MAE from perpendicular to in-plane configurations, with the sign reversal of MAE permitting voltage-controlled stabilization of topological spin textures across a well-defined biasing regime. Our results provide a platform for understanding topological magnetic quasiparticles and their interactions in two-dimensional systems. This may contribute to elucidating the mechanisms of the BKT transition and related fundamental quantum phenomena such as superfluidity and superconductivity[6, 7]. Besides, the electric-field-controlled creation and annihilation of skyrmions break through the limitations of the energy-intensive current-driven paradigm in racetrack memory[23-25], which may revolutionize next-generation memory devices and information technologies.

The scalar spin chirality of topological magnetic structures has manifested itself as an emergent magnetic field ($B_e$)[14], which gives rise to a transversal electrical response (Fig. 1a) known as the conventional topological Hall effect (THE) at zero-frequency limits[26, 27]. The emergent magnetic field arising from skyrmion potentially gives rise to the topological magneto-optical effect[28, 29], which can be considered as the optical-frequency counterpart of the conventional THE (Fig. 1b). Tunable optical-frequency permit the identification of topological magnetic structures in diverse material systems—including metals, semiconductors and insulators—thereby extending the utility of the topological magneto-optical effect beyond the metallic regimes accessible to conventional THE. To investigate magnetoelectric coupling in the ideal 2D limit, we engineered monolayer $CrI_3$ devices supporting both perpendicular electric fields and electrostatic doping. Monolayer $CrI_3$ was exfoliated mechanically from bulk crystals onto a hBN flake on 285-nm-thick $SiO_2$/Si substrates in a nitrogen-filled glove box (see Methods). A micrograph of the monolayer $CrI_3$ and the device is shown in Extended Data Fig. 1a-b. Extended Data Fig. 1c illustrates the structure of a representative monolayer $CrI_3$ device. This vertically stacked configuration consists of a monolayer $CrI_3$, hBN flakes and graphite contacts, in which the hBN flakes act as the dielectric layers for providing electrostatic doping and electric field. The dielectric layer of hBN has a thickness of 9.8 nm (Extended Data Fig. 2) and a dielectric constant of 2.9 (Extended Data Fig. 3; see Methods), consistent with the previous reports[30, 31]. Reflective magnetic circular dichroism (RMCD) measurements were performed on a designated region of the device, using a laser with a wavelength of 633 nm and a fixed power of 3 μW (see Methods). Figure 1c shows a representative RMCD signal of the monolayer $CrI_3$ device as a function of the out-of-plane magnetic field ($\mu_0 H$) at zero gate voltage, with arrows indicating the direction of the magnetic field sweep. The results of monolayer $CrI_3$ are in full agreement with the previous reports[1]. Namely, monolayer $CrI_3$ exhibits ferromagnetism with a coercive field ($\mu_0 H_c$) of approximately 0.04 T and a Curie temperature ($T_c$) close to 40 K (Extended Data Fig. 4). And the absence of a high-field step in the RMCD loop (Extended Data Fig. 5a), excludes interlayer antiferromagnetic coupling and validates the monolayer[32]. Furthermore, the polarization-resolved and temperature-dependent Raman measurements of spin-lattice coupling further confirm the monolayer feature[33, 34], in which the $A_{1g}$ phonons are only detected at 127.8 $cm^{-1}$ in cross- (XY) and co-linear (XX) polarization configurations and significantly increased in XY configuration below 40 K. This temperature-dependent evolution of the polarization-resolved Raman features and the RMCD signals provide explicit evidence that the CrI$_3$ sample is a monolayer (Supplementary Note 1, Extended Data Fig. 4 and 5).

Figure 1d shows the RMCD signal as a function of the out-of-plane magnetic field under an applied voltage of 8 V, revealing a pair of anomalous RMCD peaks near the spin-switching transition field. The polar RMCD signal intensity ($I_{RMCD}$) is approximately proportional to $\sigma_{xy}(\omega)$, as expressed by the following equation[35-37]

$$I_{RMCD} = \frac{2Z_0 d\sigma_{xy}(\omega)}{1-(n_S + Z_0 d\sigma_{xx}(\omega))^2} \quad (1)$$

Here, $Z_0$ is the impedance of free space, $n_s$ denotes the refractive index of the substrate, and $d$ represents the thickness of the monolayer CrI$_3$, approximately 0.7 nm; $\sigma_{xx}(\omega)$ and $\sigma_{xy}(\omega)$ are the longitudinal and transverse components of the optical Hall conductivity, respectively. Analogous to the analytical framework for the conventional topological Hall effect, $\sigma_{xy}(\omega)$ can be separated into two primary components: one originating from ferromagnetism, denoted as $\sigma_{xy}^A(\omega)$, and the other arising from topological spin textures, denoted as $\sigma_{xy}^T(\omega)$. The $\sigma_{xy}^A(\omega)$ is proportional to the magnetization ($M$), while $\sigma_{xy}^T(\omega)$ is proportional to the emergent magnetic field $\boldsymbol{B}_e$ induced by topological spin textures[29], which can be determined by evaluating the solid angle traced by the spin moment $\boldsymbol{n}$ as it winds within the $xy$-plane[15], given by

$$\sigma_{xy}^T(\omega) \propto \boldsymbol{B}_e = \frac{1}{2}\boldsymbol{n}\cdot\left(\frac{\partial \boldsymbol{n}}{\partial x} \times \frac{\partial \boldsymbol{n}}{\partial y}\right)\hat{\boldsymbol{z}} \quad (2)$$

For a skyrmion, $\boldsymbol{B}_e$ is quantized: the integral over the 2D space is simply $2\pi$ times the topological charge ($Q$) of skyrmions, $Q = \frac{1}{4\pi}\int \boldsymbol{n}\cdot\left(\frac{\partial \boldsymbol{n}}{\partial x} \times \frac{\partial \boldsymbol{n}}{\partial y}\right)dxdy$. Therefore, spin textures with non-zero $Q$ can induce differences in the reflectivity between the left- and right-handed circularly polarized lights, even in the absence of a net magnetic moment. We refer to this effect as reflective topological circular dichroism (RCD$_T$), which can also lead to the rotation of the polarization direction of linearly polarized reflected light, known as the topological Kerr effect. The RMCD loops illustrated in Fig. 1e can be expressed as the sum of two components,

$$I_{RMCD} = I_{RCD(M)} + I_{RCD(T)} \quad (3)$$

Here, $I_{RCD(M)}$ is the intensity of conventional reflective circular dichroism (RCD$_M$), which is proportional to the out-of-plane net magnetization of the ferromagnetic background (highlighted in purple, see Figs. 1c-f), while $I_{RCD(T)}$ is the intensity of $RCD_T$ and proportional to the density of $Q$ (highlighted in orange, Figs. 1d-f). The upward and downward RMCD peaks suggest the generation of the topologically nontrivial skyrmions with $+Q$ and $-Q$, respectively[15, 16], in the vicinity of the spin-switching field[38]. Figure 1e illustrates the relationship between the magnetic field and the RMCD signal at an applied voltage of 12 V. The residual signal of $RCD_M$ hysteresis is significantly reduced, and the saturation magnetic field is increased, suggesting a decrease in the perpendicular magnetic anisotropy (PMA)[39]. Meanwhile, $RCD_T$ becomes more pronounced due to the increased density of $Q$. Remarkably, under a voltage of 16 V, the $RCD_M$ hysteresis completely vanishes, leaving only a significant $RCD_T$, accompanied by a giant topological circular dichroism (Fig. 1f), akin to the giant topological Hall effect observed in skyrmion lattices[15, 16]. This indicates that the perpendicular anisotropy is lost, and due to electron doping[39, 40], the system transitions to a small in-plane anisotropy, while the density of skyrmions significantly increases with the assistance of an out-of-plane magnetic field.

Figure 2a presents the RMCD intensity as a function of gate voltage ($V_g$) and magnetic field ($\mu_0 H$), sweeping from positive to negative values. In this magnetic phase diagram, white regions indicate near-zero RMCD intensity, corresponding to an in-plane ferromagnetic state. Darker red and blue areas represent two out-of-plane ferromagnetic states with strong RMCD signals of opposite signs. Light blue and red regions denote canted ferromagnetic states with opposite out-of-plane components. Notably, a distinct red stripe appears near the critical magnetic field for the spin-up to spin-down transition, identified as a topological magnetic phase with positive topological charge. In the non-topological phase region, increasing voltage drives a smooth transition from out-of-plane to in-plane ferromagnetism, suggesting a voltage-induced evolution of the perpendicular magnetic anisotropy toward in-plane anisotropy (highlighted by the blue arrows)[40]. The corresponding phase diagram, obtained by sweeping the magnetic field from negative to positive values, is shown in Extended Data Fig. 6a. Similar to Fig. 2a, a distinct blue stripe appears near the critical magnetic field for the spin-down to spin-up transition, representing a topological magnetic phase with negative topological charge. Figure 2b displays the intensities of the topological circular dichroism $RCD_T$ as a function of $V_g$ and $\mu_0 H$, revealing a critical voltage for the emergence of the topological magnetic orders near 5 V. Additionally, the critical magnetic field distinguishing the ferromagnetic and topological magnetic phases varies with the applied voltage. $RCD_T$ is obtained by subtracting $RCD_M$ from the total RMCD, as described in Equation (3) and Fig. 1e. Figure 2c shows the $RCD_T$ loops measured at several representative voltages, extracted from Fig. 2b. As the voltage increases, the $RCD_T$ peaks with positive and negative topological charges shift linearly toward higher magnetic fields, with slopes of 22 Oe/V and 13 Oe/V, respectively (Fig. 2d). This linear shift demonstrates precise and continuous voltage-controlled tuning of the topological magnetic phases. This behavior arises from the reduction of the perpendicular anisotropy induced by higher voltages, thus necessitating a stronger out-of-plane magnetic field to stabilize the skyrmions. However, above 14 V, the positions of the $RCD_T$ peaks abruptly shift toward lower magnetic fields. It suggests a transition from the perpendicular magnetic anisotropy to in-plane one, thereby causing a sudden shift in the magnetic field required to stabilize the topological structure. Figure 2e illustrates the electrically controlled $RCD_T$ under three selected magnetic fields with opposite directions, extracted from Fig. 2b. The $RCD_T$ signal remains near zero at voltages below approximately 5 V, corresponding to the ferromagnetic ordering. As the voltage increases to a critical value around 5 V, the $RCD_T$ signal rises sharply, indicating an electrical switch from trivial ferromagnetic to topological magnetic states. With voltage further increasing, the $RCD_T$ signal gradually rises and saturates, suggesting an increase in skyrmion density. Near 14 V, the $RCD_T$ signal undergoes a sudden change, suggesting that a fundamental shift in the magnetic anisotropy may trigger the nucleation of the skyrmions with new characteristics, thereby causing a sharp variation in the $RCD_T$ signal. We refer to the skyrmions induced by an out-of-plane magnetic field under perpendicular anisotropy as Type-I skyrmions, and those formed under a small in-plane magnetic anisotropy induced by an out-of-plane magnetic field as Type-II skyrmions (Fig. 2a and 2e). This abrupt behavior at 14 V is consistent with the voltage-induced shifts observed in $RCD_T$ peak positions (Fig. 2d) and intensities (Extended Data Fig. 6b). Furthermore, the increasing voltage results in a corresponding rise in $RCD_T$, attributed to an increase in the density of type-II skyrmions. The voltage can continuously regulate the density and morphology of the topological magnetic quasiparticles, and one of the key reasons is that the voltage enables continuous adjustment of the magnetic anisotropy energy, even changing its sign[40, 41], leading to a gradual transition from perpendicular anisotropy to in-plane anisotropy (Extended Data Fig. 7c).

The stability of the skyrmions is governed by two critical parameters: the DMI and magnetic anisotropy, both originating from the fundamental mechanism of the spin–orbit coupling[41-43]. As illustrated in Extended Data Fig. 8a, the application of an out-of-plane electric field in monolayer CrI$_3$ breaks spatial inversion symmetry, leading to the emergence of a significant DMI[44]. The magnetic anisotropy energy (MAE) in CrI$_3$ monolayers is intimately associated with the spin and orbital structures near the Fermi level. First-principles calculations have demonstrated that the MAE is primarily governed by the indirect spin-orbit coupling (SOC), with the direct SOC contribution being negligible[45]. By the perturbation theory, the MAE can be succinctly expressed as[46]

$$MAE = \xi^2 \sum_{\mu,o,\sigma,\sigma'} \sigma\sigma' \frac{\left|\langle o,\sigma|L_z^I|\mu,\sigma'\rangle\right|^2 - \left|\langle o,\sigma|L_x^I|\mu,\sigma'\rangle\right|^2}{E_{\mu,\sigma} - E_{o,\sigma'}} \quad (4)$$

Here, $u$ and $o$ denote the unoccupied and occupied states, respectively, $E_{u/o,\sigma}$ represents the band energy of these states, and the spin indices $\sigma/\sigma'$ range over $\pm 1$, corresponding to the two orthogonal spin states at the $k$-point. Previous theoretical investigations have established that $\left|\langle o,\sigma|L_z^I|u,\sigma'\rangle\right|^2 - \left|\langle o,\sigma|L_x^I|u,\sigma'\rangle\right|^2$ is negative[47]. In addition to the contribution from the orbital angular momentum, the relative spin polarization ($\sigma$, $\sigma'$) of the states near the Fermi level must also be considered. When CrI$_3$ is slightly n-doped, the two lowest conduction bands act as the unoccupied and occupied states, with their parallel spin alignment ($\sigma\sigma'=1$, right panel in Extended Data Fig. 8b) resulting in a negative MAE, which favors an in-plane magnetic anisotropy. In stark contrast, under slight p-doping, the dominant contributions originate from the two highest valence bands with oppositely aligned spins ($\sigma\sigma'=-1$, left panel in Extended Data Fig. 8b), resulting in a positive MAE that favors perpendicular magnetic anisotropy. In our experimental observations, the application of a positive voltage results in electron doping and further induces a transition from perpendicular magnetic anisotropy to in-plane anisotropy in monolayer CrI$_3$, while a negative voltage (hole doping) strengthens the perpendicular out-of-plane magnetization (Extended Data Fig. 7). Theoretically, MAE ($K$) must satisfy $K \leq \pi^2 \cdot D^2 / 16 \cdot J$ for enabling the formation of a skyrmion lattice[41-43], implying that the skyrmions are stabilized only within a regime of weak anisotropy. Here, $D$ denotes the Dzyaloshinskii–Moriya interaction energy, while $J$ represents the exchange interaction energy. Extended Data Fig. 8c illustrates the relationship between the factor $\eta=\pi^2 \cdot D^2/16 \cdot J$, MAE ($K$), and gate voltage, where the $D$ term is induced by the electric field and the $K$ term arises from electrostatic doping[40, 44]. Using a simple parallel-plate capacitor model[48, 49], the corresponding voltage values can be inferred from the electric field strength and doping concentration within the CrI$_3$ monolayer (see Methods). This approach enables a comparative analysis for the stabilization of the skyrmions between the voltage range predicted by theory and the experimentally observed voltage window. The theoretically predicted voltage thresholds for the type-I skyrmion formation ($0 < K < \eta$, indicated by red dashed lines in Extended Data Fig. 8c) and the transition to the type-II skyrmions ($K < 0$, indicated by blue dashed lines) closely correspond to the experimental voltages at which the type-I skyrmion remain stable (highlighted in light yellow area) and at which the type-II skyrmions are stable (highlighted in light purple area).

The evolution of the magnetic structures in monolayer $CrI_3$ under out-of-plane magnetic fields at different voltages was simulated by solving the Landau-Lifshitz-Gilbert (LLG) equation (see Methods). The parameters for atomic-scale spin dynamics simulations were selected based on the criteria for the skyrmion formation as illustrated in Extended Data Fig. 8c. Figure 3a presents the micromagnetic simulation results at a voltage of 16 V (in-plane anisotropy, $K < 0$), depicting the evolution of the magnetic domains as the out-of-plane magnetic field $B/J$ is swept from +0.214 to -0.214. As the positive magnetic field decreases to zero, the system transitions from a magnetized-up single-domain ferromagnetic state to the stripe domains. With the increase of the negative magnetic field, the stripe domains begin to fragment, forming the isolated skyrmions with a positive topological charge. Ultimately, under a stronger negative magnetic field, the skyrmions are annihilated, and the system transitions to a magnetized-down ferromagnetic state. Conversely, when the magnetic field is swept from -0.214 to +0.214 (Fig. 3b), the system evolves from a magnetized-down ferromagnetic state to stripe domains. As the positive magnetic field increases, the stripe domains break apart, forming the isolated skyrmions with a negative topological charge. Under a stronger magnetic field, the skyrmions gradually shrink and are eventually destroyed, reverting the system to a magnetized-up ferromagnetic state. As shown in Figure 3c, the variation of $M_z$ with the magnetic field exhibits no hysteresis, which is attributed to the in-plane anisotropy ($K < 0$) induced by high electron doping[40]. Figure 3d illustrates the variation of the topological charge density with the magnetic field. When the magnetic field is swept from positive to negative, a positive topological peak appears at the negative critical magnetic field. Conversely, when the magnetic field is swept in the reverse direction, a negative topological peak emerges at the positive critical magnetic field, consistent with the variation of $RMD_T$ with the magnetic field (Fig. 1f). The inset in Fig. 3d illustrates the spin distribution of a single skyrmion in the magnetic field near the topological peak (the arrows denote the direction of the spins, and the color indicates the magnitude of $M_z$, consistent with the color bar in Fig. 3a), confirming that the skyrmion is of the Néel type. Figure 3c schematically demonstrates the variation of the sum of $M_z$ and the topological charge with the out-of-plane magnetic field, which aligns with the experimentally observed RMCD loop featuring anomalous peaks and no residual signal of $RCD_M$ (Fig. 1f). This confirms the core conclusion: the total RMCD signal intensity $I_{RMCD}$ can be expressed as Eq.3 contributing to the net out-of-plane magnetization $M_z$, and the density of the topological charge $Q$. Extended Data Figs. 9a and 9b illustrate the evolution of the magnetic structure under an out-of-plane magnetic field at a voltage of 8 V (perpendicular magnetic anisotropy, $K > 0$). The evolution behavior of the magnetic structure is similar to that observed at 16 V. However, due to the perpendicular anisotropy anisotropy, the $M_z$ loop exhibits a distinct opening, indicating non-zero remanent magnetization (Extended Data Fig. 9c), which is consistent with the non-zero remanent RMCD signal observed around 8 V (Fig. 1d). Additionally, unlike the in-plane anisotropy at 16 V, the density of the skyrmions formed under the perpendicular anisotropy anisotropy at 8 V is lower, manifested as a smaller intensity of the topological peak (Fig. 3d and Extended Data Fig. 9d). This behavior aligns with the voltage-dependent variation of $RCD_T$ peak (Fig. 2c and Extended Data Fig. 6b).

To further explore the properties of the voltage-induced two-dimensional topological magnetic phase, Figure 4a displays the RMCD intensities as a function of the magnetic field at various temperatures under a 16 V voltage. As the temperature increases, the topological magnetic phase undergoes a transition to a in-plane ferromagnetic state at approximately 24 K, followed by a transition from the in-plane ferromagnetism to a paramagnetic state around 27 K. The Curie temperature ($T_c$) of monolayer $CrI_3$ under a 16 V voltage is lower than that of intrinsic monolayer $CrI_3$ (Extended Data Fig. 4) because the electron doping reduces the MAE (Fig. 3c). On the contrary, applying a voltage of -14 V slightly increases the $T_c$ of $CrI_3$ (Extended Data Fig. 10) due to the enhancement of MAE through the hole doping[40]. This behavior is attributed to the fact that the $T_c$ of monolayer $CrI_3$ is primarily governed by magnetic anisotropy induced by the SOC effects[47]. Figures 4b and 4c show the phase diagrams of topological magnetic orders in monolayer $CrI_3$ as functions of temperature and magnetic field. The results indicate that the magnetic field range for stabilizing the topological phase is narrowed as the temperature increases. Additionally, the magnetic fields corresponding to the maximum density of the topological quasiparticles—the positions of the RMCD peaks with positive topological charge (upward peaks) and negative topological charge (downward peaks)—decrease linearly with temperature, exhibiting slopes of 11.5 Oe/K and 10.4 Oe/K, respectively (Figure 4d). Furthermore, the half-peak width of the RMCD peaks, which characterizes the robustness of topological quasiparticles against magnetic fields, also gradually diminishes as the temperature rises. This behavior closely resembles the field-temperature phase diagrams of the skyrmion lattices in bulk

$Gd_2PdSi_3$ and $GdRu_2Ge_3$[15, 16]. Figure 4e illustrates the critical temperature behavior of the topological phase transition by comparing the temperature dependence of RMCD under different magnetic fields (horizontal cuts of the phase diagrams in Figs. 4b and 4c). The critical temperature shifts from approximately 15 K at ±0.04 T to 24 K at ±0.02 T. This indicates that as the out-of-plane magnetic field increases, the critical temperature decreases. The evolution of the topological magnetic phase induced by a 14 V voltage, as a function of temperature and magnetic field, is similar to that observed at 16 V (Extended Data Figs. 11 and 12). However, compared to the 16 V case, a slight perpendicular magnetic anisotropy is persisted. Therefore, in addition to the significant $RCD_T$ peaks, an open $RCD_M$ loop was also observed (detailed discussion is provided in Supplementary Notes 2).

In conclusion, we demonstrated a voltage-control of the topologically protected spin textures in monolayer $CrI_3$. By utilizing the voltage-induced synergistic effects between DMI and MAE, we achieved the generation and continuous manipulation of the skyrmion quasiparticles at the two-dimensional limit, providing a powerful tool for controlling the topological magnetic states at a single atomic layer limit. The results highlight the potential of the voltage-driven modulation of the multiple spin-orbit interactions in creating a diverse range of two-dimensional magnetic textures with distinct topological properties. The precise electric-field control in engineering topological magnetic quasiparticles opens up new possibilities for ultra-high-density, ultra-low-power, and high-speed spintronic applications, while also broadening the scope of topological magneto-optical effects, thereby advancing the development of cutting-edge magneto-optical information storage and processing technologies.

## Methods

### Sample fabrication

Monolayer $CrI_3$ was mechanically exfoliated from bulk crystals using PDMS films in an inert glovebox. The bulk $CrI_3$ crystals were synthesized via chemical vapor transport from elemental precursors with a Cr molar ratio of 1:3. The exfoliated flakes of hBN, $CrI_3$ and graphene were sequentially transferred onto pre-patterned Au electrodes on $SiO_2$/Si substrates to form heterostructures. These samples were then in-situ loaded into an optical cryostat within the glovebox for magneto-optical-electric joint measurements. Throughout the fabrication and measurement processes, the $CrI_3$ samples were kept isolated from the atmosphere.

### Magneto-optical measurement

Polar RMCD and Raman measurements were conducted using a magneto-optical-electric joint-measurement scanning imaging system (MOEJSI)[50]. This system is based on a Witec Alpha 300R Plus low-wavenumber confocal Raman microscope, integrated with a closed-cycle superconducting magnet (7 T) and a cryogen-free optical cryostat (10 K), which features a specially designed sample mount and electronic transport measurement setup.

Raman signals were recorded using the Witec Alpha 300R Plus, coupled with a closed-cycle He optical cryostat and a superconducting magnet. A 50× objective (NA = 0.45) was employed for measurements at 10 K and under magnetic field conditions. The signals were collected via a photonic crystal fiber and directed into the spectrometer with an 1800 g/mm grating. Polarization-resolved Raman spectra were obtained by rotating the analyzer placed before the fiber. The excitation laser at 633 nm (1.96 eV) was set to an intensity of approximately 0.3 mW, with a typical integration time of 120 s.

For polar RMCD measurements, a 633 nm laser (~3 µW) modulated by a photoelastic modulator (PEM, 50 kHz) was reflected by a non-polarizing beamsplitter (R/T = 30/70) and focused onto the sample using a 50× objective (NA = 0.55, Zeiss). The reflected beam, collected by the same objective, passed through the beamsplitter and was detected by a photomultiplier tube (PMT), coupled to a lock-in amplifier, the Witec scanning imaging system, a superconducting magnet, and a voltage source meter.

**Estimation of the electric field strength and doping concentration of CrI$_3$ in the device under the voltage**

We evaluate the doping level and electric field strength of monolayer CrI$_3$ in our devices under electrostatic gating using a parallel-plate capacitor model. A simple heterostructure has been employed in this study: graphene–hBN–CrI$_3$–graphene (Fig. 1c). The doping density of CrI$_3$ ($n_{Cr}$) and the electric field ($E_{Cr}$) of CrI$_3$ are obtained under the applied voltage ($V$) between the two graphene electrodes.

$$E_{Cr} \approx \frac{V}{d_{Cr} + \frac{\varepsilon_{Cr} \cdot d_{BN}}{\varepsilon_{BN}}} \quad (4)$$

$$n_{Cr} \approx \frac{V}{4\pi k \left( \frac{d_{Cr}}{\varepsilon_{Cr}} + \frac{d_{BN}}{\varepsilon_{BN}} \right)} \quad (5)$$

Here, $d_{Cr}$ and $d_{BN}$ represent the thicknesses of CrI$_3$ and hBN, respectively, $\varepsilon_{Cr}$ and $\varepsilon_{BN}$

are the dielectric constants of CrI$_3$ and hBN, and $k$ is the electrostatic constant. The dielectric constant of hBN is measured to be 2.9 using a ferroelectric analyzer[50], while the dielectric constant of CrI$_3$ is 7[51]. The thickness of hBN is determined to be 9.8 nm using optical contrast methods[52], while the thickness of a monolayer of CrI$_3$ is 0.7 nm. By calculating equations (4) and (5), along with the theoretical predictions of the electric field–DMI function[44] and the doping concentration–magnetic anisotropy function[40], we are able to compare the experimentally observed voltage window for the emergence of topological magnetic order with the theoretically predicted critical voltage for its onset (Extended Data Fig. 8c).

**Optical contrast measurement**

White light is normally incident on the hBN-SiO$_2$-Si structure and focused into a small spot using a 50×, NA=0.45 microscope objective. The reflected light is then collected, passes through a beamsplitter, and is directed onto the entrance slit of a spectrometer. The spectrally resolved reflection signal is captured by a CCD at the spectrometer's exit aperture. In our experiments, we measure the optical contrast of the three-layer structure, which is defined as:

$$C = \frac{R_{SiO2} - R_{SiO2+hBN}}{R_{SiO2}} \quad (6)$$

Where $R_{SiO2}$ represents the reflection coefficient at normal incidence for a bare SiO$_2$/Si substrate, while $R_{SiO2+hBN}$ corresponds to the reflection coefficient for a substrate covered with hBN. This defines the normalized change in reflectivity of the hBN layer relative to the underlying substrate[52].

**Atomic-scale spin dynamics simulations**

The numerical simulation were performed by solving the atomic-scale LLG equation $-\frac{1+\alpha^2}{\gamma}\frac{d\bm{m}_i}{dt} = \bm{m}_i \times \bm{H}_i + \alpha \bm{m}_i \times (\bm{m}_i \times \bm{H}_i)$ with periodic boundary condition, where the $\bm{m}_i$ is the reduced magnetization of the $i$-th Cr atom, $\gamma = 1.76 \times 10^{11}$ T$^{-1}$s$^{-1}$ is the gyromagnetic ratio, $\alpha$ is the Gilbert damping constant (set to 1.0 to significantly reduce the evolution time) and $\bm{H}_i = -\frac{\delta \mathcal{H}}{\mu_s \delta \bm{m}_i}$ is the effective field with the magnetic moment $\mu_s$ = 2.95 $\mu_B$, respectively. The Hamiltonian $\mathcal{H}$ of CrI$_3$ monolayer reads

$$\mathcal{H} = \frac{S^2}{2} \sum_{\langle i,j \rangle} \left( J \boldsymbol{m}_i \cdot \boldsymbol{m}_j + \boldsymbol{D} \cdot \boldsymbol{m}_i \times \boldsymbol{m}_j \right) - \frac{KS^2}{2} \sum_i (\boldsymbol{m}_i \cdot \mathbf{z})^2 - \sum_i \boldsymbol{m}_i \cdot \mathbf{B}$$

, where $S = 3/2$ is the magnitude of spin, the $J$ is Heisenberg exchange parameter, $\boldsymbol{D}$ is the Dzyaloshinskii-Moriya interaction (DMI) vector with both the in-plane component ($D_{xy}$) and out-plane component ($D_z$) to stabilize the skyrmion, $K$ is the anisotropy parameter and $\boldsymbol{B}$ is the external magnetic field, respectively. Noteworthily, the in-plane component of the DMI vector is perpendicular to the line between the two Cr atoms. In the case of perpendicular magnetic anisotropy (8 V), $J$ is set to -1.962 meV for ferromagnetic coupling, $D_{xy} = -0.189J$, $D_z = -0.119J$ and $K = 0.019J$. As for the case of in-plane anisotropy (16 V), the $J$ is set to -1.992 meV, $D_{xy} = -0.210J$, $D_z = -0.132J$ and $K = -0.013J$. The aforementioned parameters were referenced from previous studies[40, 44], and based on the stable voltage conditions for skyrmion formation as shown in Extended Data Fig. 8c, 8 V and 16 V were selected as the research parameters. Moreover, the thermal fluctuation is introduced with a stochastic field $\boldsymbol{H}_T = \boldsymbol{\eta}(\text{step}) \sqrt{2 k_B T \alpha / (\gamma \mu_s \Delta t)}$, where $\boldsymbol{\eta}(\text{step})$ is a random vector from a standard normal distribution, $k_B$ is the Boltzmann constant, $T = 2$ K is the temperature and $\Delta t = 10$ fs is the time interval. In order to reach the equilibrate state, we relax 1 ns at each magnetic field of the hysteresis loop.

## Data availability

The data that support the findings of this study are available from the corresponding authors upon reasonable request. Source data are provided with this paper.

## Acknowledgments

B.P. and L.D. acknowledge support from National Science Foundation of China (52021001). B.P. acknowledges support from the Sichuan Provincial Outstanding Youth Science Foundation Project (2025NSFJQ0018) and National Science Foundation of China (62250073, 62450003). H.L. acknowledges support from National Science Foundation of China (51972046). L.D. acknowledges support from Sichuan Provincial Science and Technology Department (Grant No. 99203070). P.Y. was supported by National Key R&D Program under Contract No. 2022YFA1402802 and the National Natural Science Foundation of China (NSFC) (Grants No. 12374103 and

## Author contributions

## Competing interests

The authors declare no competing interests.

## Additional information

**Supplementary information** is available for this paper at xxx (will be provided).

**Correspondence and requests for materials** should be addressed to Longjiang Deng, Peng Yan or Bo Peng

**Peer review information** *Nature* thanks xxx, and the other, anonymous, reviewer(s) for their contribution to the peer review of this work. A peer review file is available.

**Reprints and permission information** is available online at http://www.nature.com/reprints.

**Publisher's note** Springer Nature remains neutral with regard to jurisdictional claims in published maps and institutional affiliations.

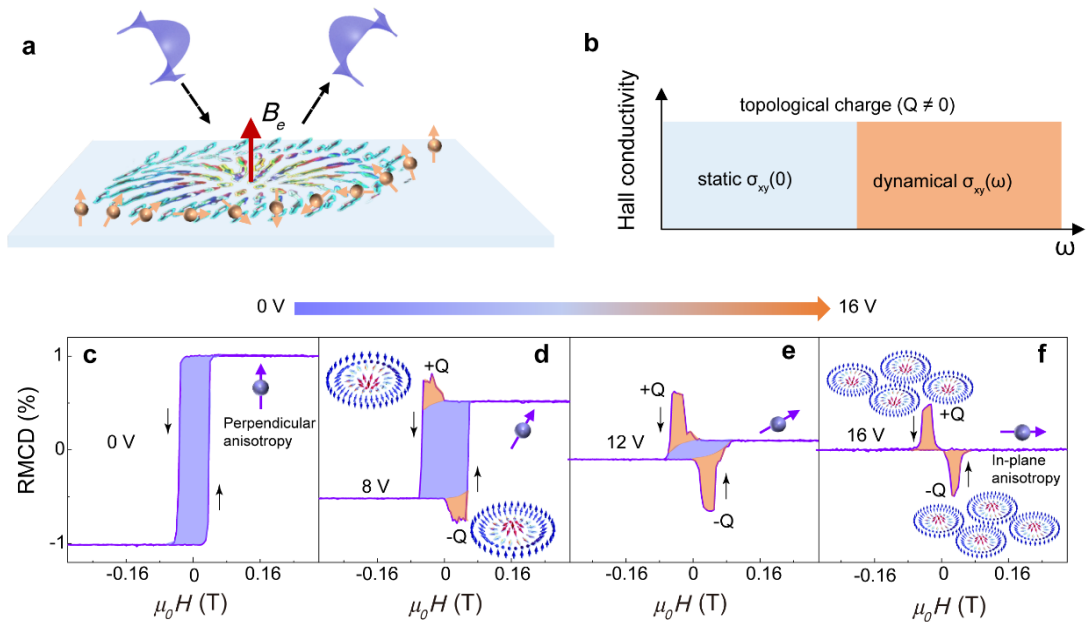

**Fig. 1 | Voltage-controlled topological circular dichroism in monolayer CrI$_3$. a,** Sketch of the topological Hall effect and topological circular dichroism induced by the emergent magnetic field. **b**, Summary of the two quantum Hall regimes. **c-f**, RMCD versus magnetic field at different gate voltages at 10 K. The black arrows indicate the direction of the field sweep, while the purple arrows with spheres schematically represent the magnetic anisotropy.

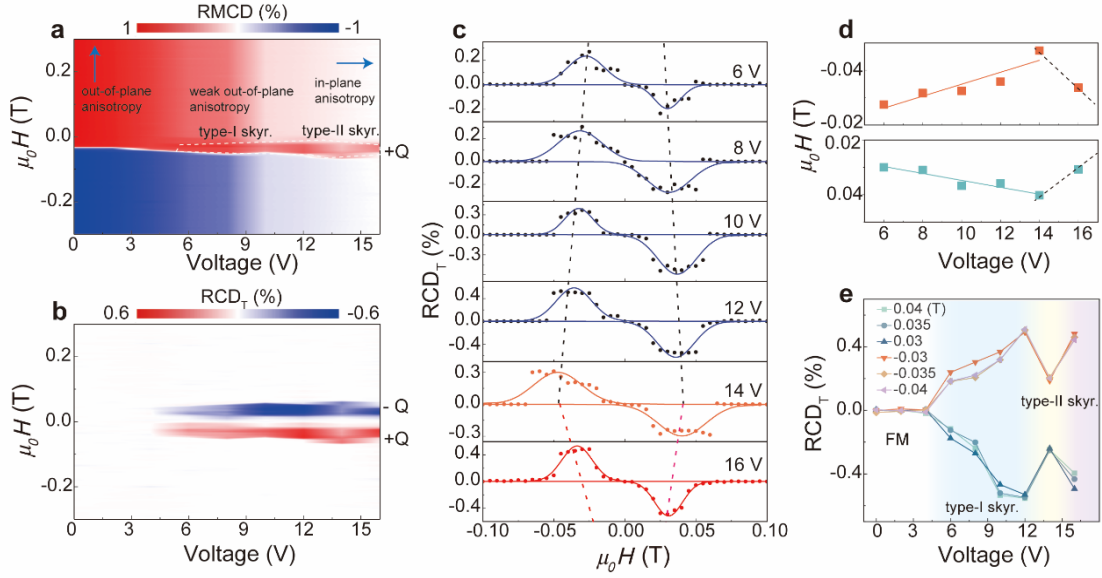

**Fig. 2 | Voltage-controlled multistage topological magnetic phase transition in monolayer CrI$_3$. a,** Intensity of the RMCD signal of a monolayer CrI$_3$ device as a function of both gate voltage and applied magnetic field (sweeping from positive to negative). The blue arrows highlight the voltage-induced evolution of the perpendicular magnetic anisotropy to the in-plane anisotropy. **b,** topological magnetic phase diagram from the RCD$_T$ as a function of gate voltage V$_g$ and external field $\mu_0H$. **c,** The RCD$_T$ loops measured at various voltages, selected from Fig. 2b. Solid dots represent the experimental values, while the solid line is derived from Gaussian fitting. **d,** The RCD$_T$ peak positions, where the density of topological quasiparticles is maximal under the corresponding magnetic fields, are plotted as a function of the applied voltage. The solid lines represent linear fits, while the dashed lines serve as visual guides. **e,** Under selected magnetic fields of $\mu_0H = \pm 0.04$ T, $\pm 0.035$ T, and $\pm 0.03$ T, the relationship between RCD$_T$ and voltage indicates a voltage-induced multistage magnetic phase transition from the ferromagnetic phase to the skyrmion phase.

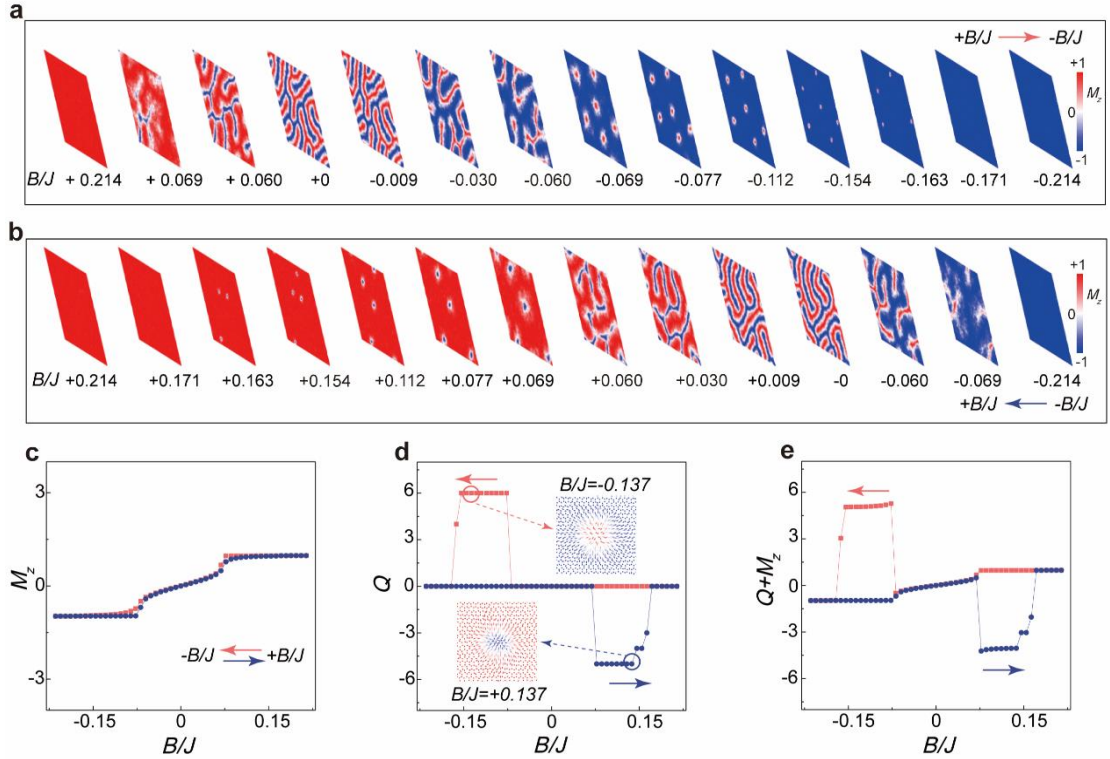

**Fig. 3 | Simulated evolution of the magnetic structures in monolayer CrI₃ induced by a magnetic field under a voltage of 16 V (in-plane anisotropy, $K < 0$). a,** A series of representative magnetic domain configurations during the process of the magnetic field sweeping from the positive maximum to the negative maximum. A 100×100 supercell was used for the magnetic simulation of the two-dimensional spin lattice, with colors mapping the out-of-plane magnetic moment components ($M_z$). **b,** A series of representative magnetic domain configurations, as the magnetic field sweeps back from the negative maximum to the positive maximum, is consistent with the RMCD loop sweeping. **c,** The variation of $M_z$ with the magnetic field ($M_z$ normalized to its saturation value). The orange line represents the magnetic field sweeping direction from positive to negative, while the blue line indicates the opposite sweeping direction. **d,** The evolution of the topological charge (per 10,000 spin sites) along the same hysteresis loop depicted in (**c**). The inset shows an enlarged view of the spin configuration of a single skyrmion, with arrows indicating the direction of the spins and colors representing $M_z$. **e,** The sum of the topological charge and $M_z$ changes along the same hysteresis loop in (**c**) and (**d**), which resembles the RMCD loop observed experimentally under a voltage of 16 V.

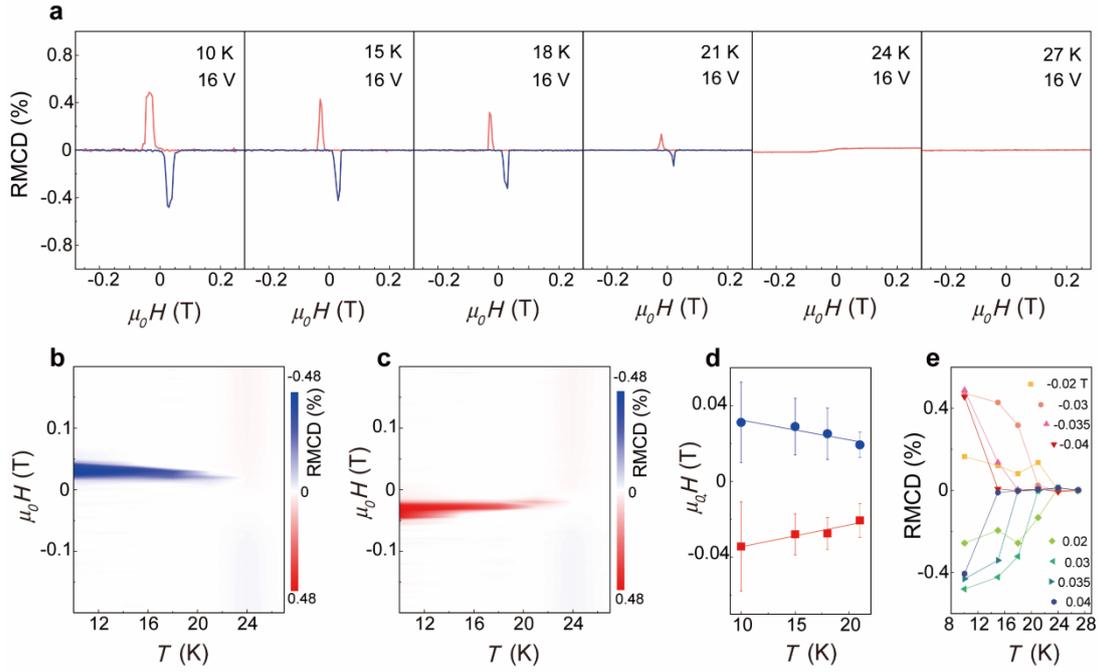

**Fig. 4 | Temperature-magnetic field phase diagram at 16 V. a,** Intensity of the RMCD signal of a monolayer $CrI_3$ device under a 16 V voltage as a function of the magnetic field at different temperatures. The orange line represents the sweeping direction from positive to negative magnetic field, while the blue line represents the sweeping direction from negative to positive magnetic field. **b,** The RMCD intensity as a function of temperature and magnetic field (sweeping from negative to positive). **c,** The time-reversal process corresponding to (**b**) (sweeping from positive to negative). **d**, The RCMD peak positions, where the density of topological quasiparticles is maximal under the corresponding magnetic fields, are plotted as a function of the temperatures. Blue solid circles and red solid squares represent the magnetic field values corresponding to the blue and red solid RMCD peaks in **a**, respectively. Solid lines denote linear fits to the experimental data, and error bars represent the half-peak widths. The peak positions and half-peak widths are obtained through Gaussian fitting. **e,** The RMCD signal intensity as a function of temperature at various positive and negative magnetic fields indicates that the out-of-plane magnetic field significantly influences the phase transition temperature from topological magnetic to ferromagnetic.